\documentclass[%
 reprint,
 amsmath,amssymb,
 aps,
]{revtex4-2}

\usepackage{graphicx}% Include figure files
\usepackage{dcolumn}% Align table columns on decimal point
\usepackage{bm}% bold math
\usepackage{caption}

\usepackage{subcaption}
\begin{document}

\preprint{APS/123-QED}

\title{Geometry-Controlled Motility of Microswimmers in elementary microfluidic confinements}% Force line breaks with \\

\author{Marvin Brun-Cosme-Bruny}
 
 \email{marvinbcb@gmail.com}
 
\author{Philippe Peyla}%
\author{Salima Rafaï}%
\affiliation{%
 Université Grenoble Alpes, CNRS, LIPhy, F-38000 Grenoble, France}%

\date{\today}% It is always \today, today,
             %  but any date may be explicitly specified

\begin{abstract}
The motility of microswimmers in confined environments is a fundamental problem in active matter physics, with direct implications for microfluidic applications and the understanding of microorganism behavior in complex natural habitats. Although the run-and-tumble dynamics of flagellated microalgae such as \textit{Chlamydomonas Reinhardtii} (CR) are well characterized in bulk suspension, the extent to which elementary geometric confinements alter their swimming remains insufficiently understood, particularly regarding the relative contributions of steric contact versus hydrodynamic interactions. Here, we experimentally investigate the trajectories of individual CR cells in a diversity of PDMS microfluidic geometries of growing complexity using single-particle tracking and statistical analysis. We show that cells in straight channels accumulate near walls and align along the channel axis, a behavior qualitatively reproduced by steric Active Brownian Particle simulations, yet showing a confinement-dependent velocity enhancement consistent with hydrodynamic wall coupling. In circular cavities with diameter below the persistence length $L_0\sim 350 $ µm, cells transition from bulk active Brownian exploration to quasi-circular wall-following trajectories. In dumbbell geometries, inter-compartment dwell length reflect purely geometric predictions, evidencing no measurable hydrodynamic contributions even for strong confinements. Together, these results demonstrate that environmental geometry can selectively amplify or suppress motility modes in active biological suspensions, opening avenues for the passive control of microswimmer transport in engineered microfluidic networks.
\end{abstract}

%Use showkeys class option if keyword
                              %display desired
\maketitle

%\tableofcontents

\section{\label{sec:level1}Introduction}

The swimming behavior of microorganisms in confined geometries lies at the intersection of active matter physics, microfluidics, and biological locomotion. Single-celled flagellated microswimmers—such as the model alga \textit{Chlamydomonas Reinhardtii} (CR)—propel themselves through fluid media by periodic flagellar beating, generating complex hydrodynamic fields that couple with the surrounding boundaries \cite{lauga2009hydrodynamics, guasto2012fluid}. Understanding how geometric confinement shapes these trajectories is of fundamental importance, not only for the biophysics of microorganism motility but also for the rational design of microfluidic devices, where the interplay between active suspension dynamics and channel geometry governs transport, dispersion, and sorting \cite{altshuler2013flow, berdakin2013influence}.

The motility of CR in bulk suspension is well described by an Active Brownian Particle model, in which directed swimming runs of characteristic persistence length $L_0 \sim 350$ µm are intermittently interrupted by stochastic reorientation events (tumbles), producing an effective diffusive behavior at long times \cite{berg1972chemotaxis, garcia2011random}. Near a planar surface, both hydrodynamic interactions and steric contact forces contribute to trajectory deflection and near-wall accumulation \cite{mino2011enhanced, berke2008hydrodynamic, spagnolie2012hydrodynamics, lagoin2025enhanced, bardfalvy2024collective}. Pusher-type swimmers, such as \textit{E. coli}, are hydrodynamically attracted to walls \cite{berke2008hydrodynamic}, while the situation for puller-type swimmers like CR is more nuanced: their far-field flow induces a repulsive interaction with walls parallel to the propulsion axis, yet near-wall contact and flagellar dynamics still drive significant surface accumulation \cite{kantsler2013ciliary}. However, contact interactions—rather than hydrodynamic attraction—are shown to be the primary mechanism for CR surface scattering.

Beyond interactions with planar walls, the motility of CR becomes even richer in more complex geometrical confinements, such as structured microenvironments containing pillars, cavities, or obstacle arrays \cite{brun2019effective, thery2021rebound, ao2014active, volpe2014simulation}. In such settings, local curvature and topological constraints profoundly modulate swimmer trajectories, leading to emergent transport behaviors distinct from bulk or near-wall motion, including trapping, guided transport, and pronounced corner accumulation in concave microchambers \cite{ostapenko2018curvature, brun2020deflection}. Our previous experimental and theoretical studies have shown that when microswimmers navigate through periodic pillar lattices or crowded media, their repeated scattering on obstacles results from a competition between quasi-deterministic deflection rules and stochastic reorientation, which in turn controls both effective diffusivity and large-scale transport anisotropy \cite{chepizhko2019ideal,alalam2022active}. Such observations reveal how geometrical disorder or periodicity can tune the transition from directed to diffusive swimming regimes, and even amplify or suppress transport depending on obstacle density and swimmer persistence. This underscores the critical role of complex topology in shaping microswimmer dispersal and accumulation within microstructured environments \cite{parra2014casimir, harris2009chlamydomonas}.

Here, we present a systematic experimental study of CR motility across a hierarchy of elementary to complex microfluidic geometries. While straight channels allow us to characterize microswimmer interactions with planar surfaces, circular cavities further reveal their behavior near concave walls. Finally, by combining both topologies, swimming trajectories are quantified in dumbbell geometries. By tracking individual cells across these configurations and analyzing swimming dynamics, we quantify the progressive modifications to swimming dynamics induced by increasing geometric complexity. Our results, supported by simulations, reveal that steric contact interactions dominate obstacle-induced reorientation for moderate confinements, while hydrodynamic coupling seems to become significant when characteristic geometric dimensions fall well below the persistence length.

\section{\label{sec:material}Materials and methods}

\subsection{Experimental set-up}
The green microalga CR is used as a model microswimmer. It is a biflagellate cell of approximately $a =10$ µm in diameter \cite{polin2009chlamydomonas}. Cells are grown in a 14 h / 10 h light/dark cycle at 22°C and harvested in the middle of the exponential growth phase. CR propels itself through fluid using its two front flagella, which beat in a breast stroke manner at a frequency $\sim~30$ Hz. In the absence of any external stimulus, the resulting swimming motion is well characterized by a persistent random walk \cite{garcia2011random}.

\begin{figure*}
\includegraphics[width=\textwidth]{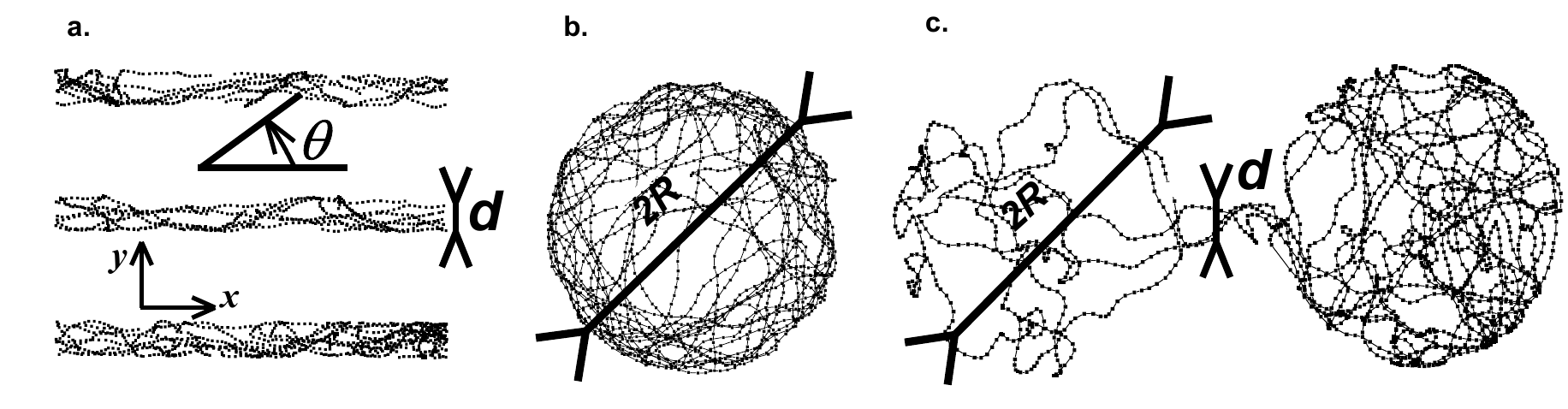}% Here is how to import EPS art
\caption{\label{merge}Map of cells’ trajectories tracked in the $xy$ plane over few minutes, within channels (a), a cavity (b) and a dumbbell (c) confinement. Time interval is 1/15 s. $d$ refers to channel width and $2R$ represents the diameter of cavities. $\theta$ is the microswimmer's orientation whose origin is the channel axis ($x$ axis).}
\end{figure*}

Complex microfluidic geometries are made in poly(dimethylsiloxane) (PDMS) using standard soft lithography processes \cite{qin2010soft}. Three elementary confinement geometries are investigated. First, straight channels of fixed height (70 µm, approximately 7 cell diameters) and varying width $d$ from 20 to 411 µm are used to study the one-dimensional lateral confinement. Second, isolated circular cavities of varying diameter $2R$, from 110 to 436 µm, provide a two-dimensional confinement geometry with curved walls. Finally, dumbbell-shaped topologies—consisting of two circular reservoirs of variable diameter $2R$ (for values 150, 210, and 266 µm) connected by a narrow channel of variable width ($d$ ranges from 30 to 130 µm) and a fixed length $\sim L_0$—enable the study of inter-compartment transport dynamics under combined confinement. Across all geometries, the characteristic length scales are chosen to span a range from below to above the persistence length $L_0$, in order to systematically probe the transition between weakly and strongly confined swimming regimes. Fig. \ref{merge} illustrates the trajectories tracked over a few minutes on cells within the different geometries, which sums up the different characteristic dimensions used throughout our study. Bovine Serum Albumin is used to coat the channel surfaces to limit cell adhesion to walls. At the beginning of each experiment, the microswimmers are observed to be homogeneously distributed within the chamber. Suspensions are used at an initial dilute volume fraction of approximately 0.05\%, ensuring that hydrodynamic pair interactions among microswimmers are negligible. For the same purpose in circular cavities and dumbbell configurations, cells are individually trapped.

Observations are performed using bright-field microscopy with an inverted microscope (Olympus IX71) coupled to a CMOS camera (Imaging Source), operated at a frame rate of 15 frames per second. The low-magnification objectives (×1.25) provide a wide field of view—up to 3614 × 2885 µm². The large depth of field associated with this low magnification ensures that swimmers are tracked throughout the full chamber height. To prevent any parasite light from triggering phototaxis, the sample is enclosed in an occluding box equipped with two red-filtered windows for visualization \cite{garcia2013light}.

Particle tracking is performed using Trackpy \cite{crocker1996methods, allan2018trackpy}, a Python library based on the Crocker and Grier algorithm. From the resulting trajectories, relevant dynamical quantities are extracted via ensemble averaging over long-duration movies (approximately 6 min), including mean-squared displacements (MSD), orientation distributions, instantaneous velocity profiles, and directional autocorrelation functions.

\subsection{Numerical simulation of Active Brownian Particle dynamics}
The particle trajectory is generated using a 2D persistent random walk model. 
At each time step $\mathrm{d}t$, the orientation angle $\theta$ is updated by 
adding a Gaussian increase in angularity $\delta\theta \sim \mathcal{N}(0,\,\sigma_\theta)$, 
where the amplitude of angular noise is related to the persistence time $t_0$ by:
\begin{equation}
    \sigma_\theta = \sqrt{\frac{\mathrm{d}t}{t_0}}
\end{equation}
The particle then moves at constant speed $v$ along the updated direction:
\begin{align}
    x_{i+1} &= x_i + v\cos(\theta_i)\,\mathrm{d}t, \\
    y_{i+1} &= y_i + v\sin(\theta_i)\,\mathrm{d}t.
\end{align}
A long persistence time $t_0$ yields a smooth directional trajectory, 
while $t_0 \to \mathrm{d}t$ recovers an isotropic diffusive random walk. 
The directional autocorrelation decays exponentially as 
$\langle \cos(\theta(t) - \theta(0)) \rangle = e^{-t/t_0}$, 
consistent with the standard persistent random walk framework.

Particle positions are stored at each frame, and the resulting trajectories 
are processed identically to experimental tracks: mean-squared displacements, 
orientation distributions, and transverse density profiles are extracted by ensemble averaging over a large number of independent realizations. 
Wall interactions are implemented as follows: upon approaching the boundary, the particle orientation 
is checked against the local obstacle. If the following trajectory step points into the obstacle, a new 
orientation is sampled with the same normal distribution until a direction pointing away from the wall is 
obtained.

\section{\label{sec:results}Results}

\subsection{\label{oneD}One-dimensional confinement: channel geometry}

\begin{figure}
\includegraphics[width=\columnwidth]{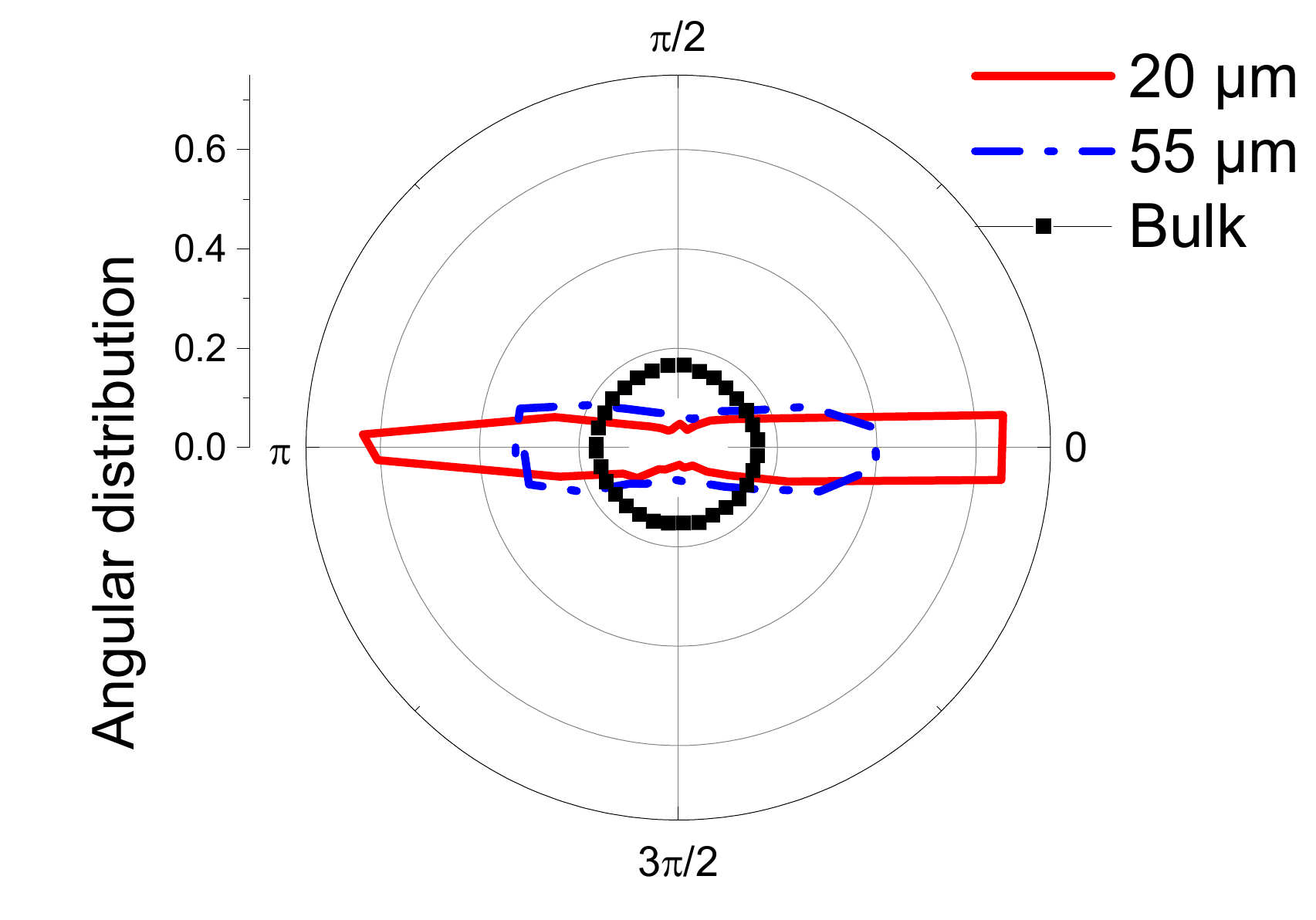}
\caption{\label{orient} Distribution of microswimmers' orientations $\theta$ for different channel widths $d$ (20 µm, 55 µm) and in bulk. Orientations are measured over a duration of 0.5 s. The channel axis, oriented horizontally, is taken as the angular origin.}
\end{figure}

We first investigate CR swimming in fixed height (70 µm) and varying width $d$, ranging from 25 µm to 410 µm, in the absence of external flow. The channel length (2 mm) is chosen large compared to $L_0$ so that cells undergo multiple runs within a single channel. The lateral confinement induces a clear orientational bias: the angular distribution of the swimming directions $\theta$, uniform in the simple medium, develops sharp peaks at 0 and $\pi$ rad—that is, along the channel axis—whose amplitude increases as $d$ decreases (fig. \ref{orient}). This alignment is a direct consequence of cell-wall interactions, which preferentially redirect cells along the channel direction.

\begin{figure}
    \centering
    % Première sous-figure
    \begin{subfigure}{\columnwidth} % Largeur de la sous-figure
        \centering
        \includegraphics[width=\textwidth]{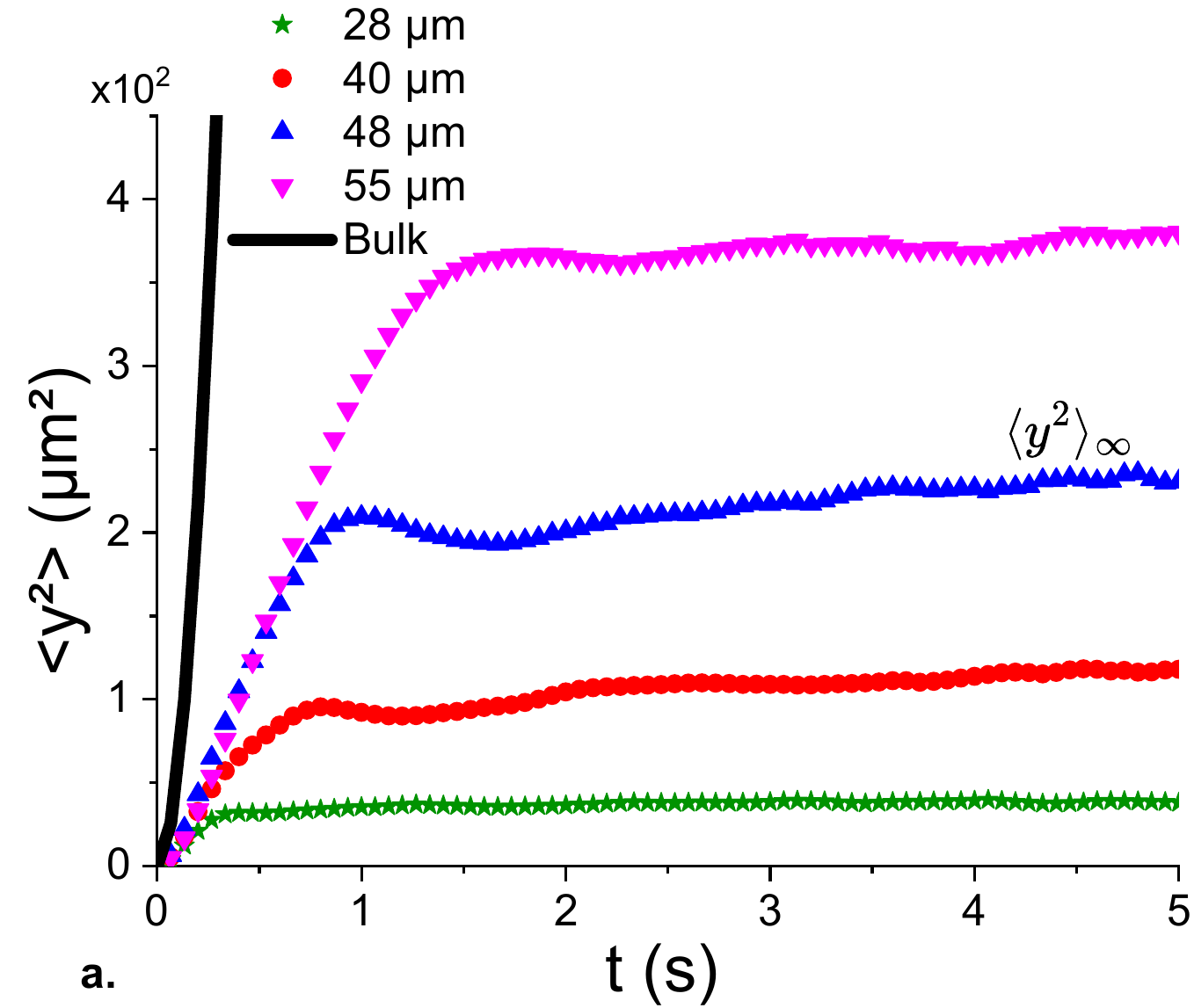}
        \label{msda}
    \end{subfigure}

    % Seconde sous-figure
    \begin{subfigure}{\columnwidth}
        \centering
        \includegraphics[width=\columnwidth]{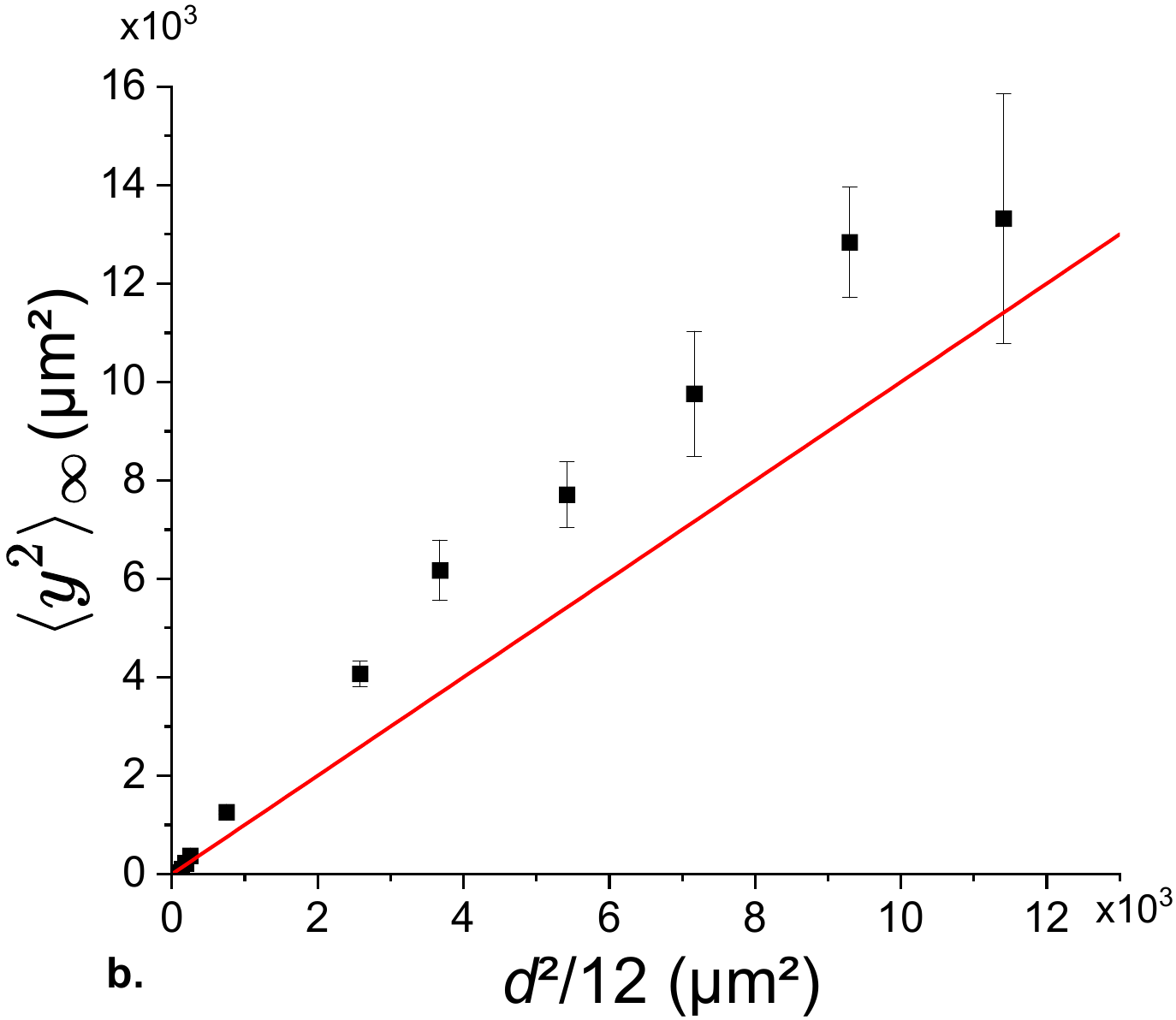}
        \label{msdb}
    \end{subfigure}
    
    \caption{a. Mean-squared displacements $\langle y^2\rangle$ projected along the $y$ axis (orthogonal to the channel axis) as a function of time, for different channel widths. b. Saturation values $\langle y^2\rangle_\infty$ of $\langle y^2\rangle$ for the different channel widths $d$, extracted from fig. 3a. The horizontal axis represents $d^2/12$, which defines the variance of a uniform distribution. The first bisector is drawn in red.}
    \label{msd}
\end{figure}

The transverse mean-squared displacement $\langle y^2\rangle(t)$ reaches a plateau $\langle y^2\rangle_\infty$, as illustrated in fig. \ref{msd}a, demonstrating the confinement of trajectories along the $y$ direction, orthogonal to the channel direction. This plateau grows with $d$ (fig. \ref{msd}b), in contrast to the long-term linear diffusive increase observed under bulk conditions \cite{garcia2011random}. Plotting $\langle y^2\rangle_\infty$ against $d^2/12$—the variance expected for a uniform transverse distribution when considering the middle of the channel width as $y=0$—reveals a systematic deviation which indicates that cells do not distribute uniformly across the channel width.

\begin{figure}
\includegraphics[width=\columnwidth]{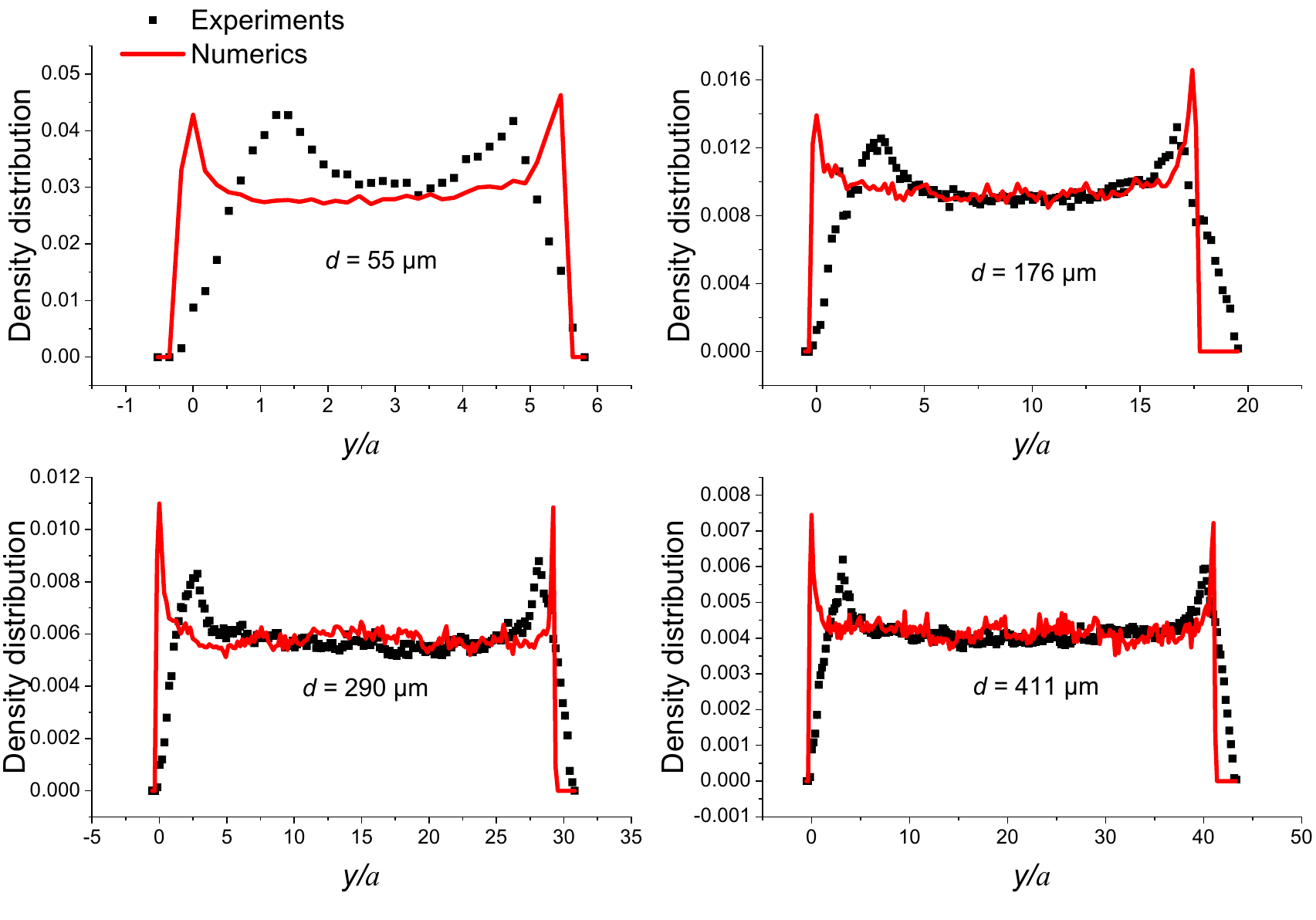}
\caption{\label{density} Transverse cell density distribution along the $y$ axis (normalized by the cell diameter $a$), for different channel widths $d$. Experimental density profiles are averaged over the longitudinal direction of the channel and over all 900 frames of the video acquisition. Numerical simulations (solid lines) model the swimmer as a pointlike particle undergoing a two-dimensional persistent active Brownian motion.}
\end{figure}

Direct measurement of the transverse density profile $\rho(y)$ confirms this in fig. \ref{density}: cells accumulate sharply near the walls, accumulating significantly more from the walls than in the channel center. This near-wall accumulation is qualitatively reproduced by a purely steric Active Brownian Particle simulation in which a pointlike swimmer bouncing off a wall randomly reorients until an outward direction is selected (see the Appendix), demonstrating that hydrodynamic interactions are not required to explain the observed density profiles.

\begin{figure}
\includegraphics[width=\columnwidth]{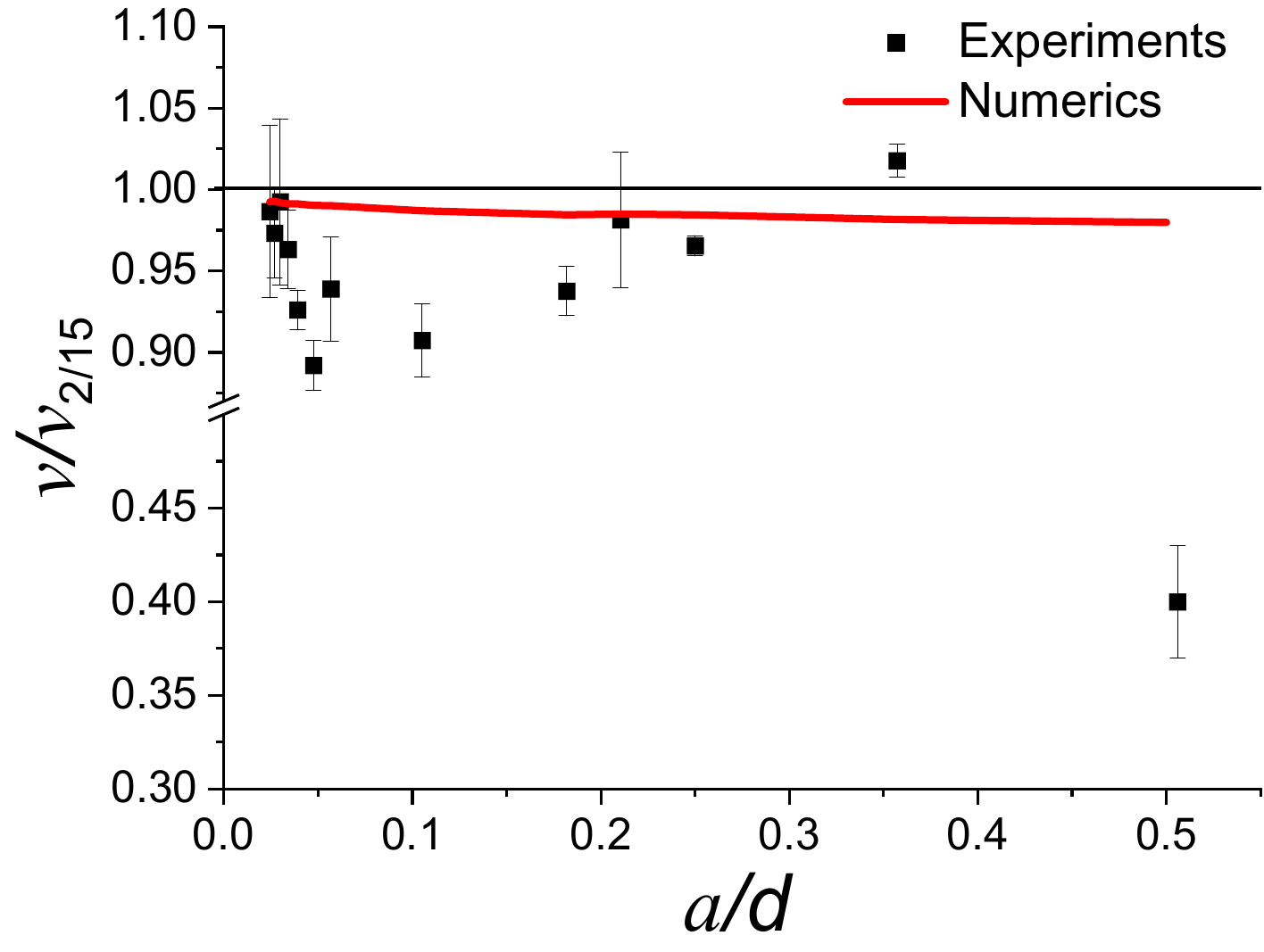}
\caption{\label{velocity} Instantaneous swimming speed $v$ of the microswimmer in a channel, normalized by the bulk reference speed measured over 2/15 s, $v_{2/15} = 130$ µm/s, as a function of the confinement ratio $a/d$, where $a = 10$ µm is the cell diameter.}
\end{figure}

Nevertheless, measurements of the instantaneous swimming speed $v$ (measured over 2/15 s) as a function of confinement ratio $a/d$ (where $a = 10$ µm is the cell diameter) reveal a non-monotonic behavior (fig. \ref{velocity}). $v$ first decreases as the confinement increases, consistent with more frequent wall collisions, then progressively recovers its bulk value at intermediate confinement, before decreasing again at extreme confinement where flagellar contact with the walls becomes significant. This improvement in velocity in intermediate confinement—where cells swim faster despite proximity to walls—is consistent with a hydrodynamic coupling between the swimmer's flow field and the confining surfaces, as previously reported for microswimmers in channels \cite{ishikawa2025transport, wu2016amoeboid}. The steric numerical simulations do not reproduce such velocity variations, further supporting the involvement of hydrodynamic effects in confined environments. Channel confinement therefore seems to combine two distinct physical contributions: steric contact interactions, which govern the orientational bias and near-wall accumulation, and hydrodynamic wall coupling, which modulates propulsion speed at strong confinement.

\subsection{\label{2D}Two-dimensional confinement: cavities and dumbbell geometries}

\begin{figure}[h]
\includegraphics[width=\columnwidth]{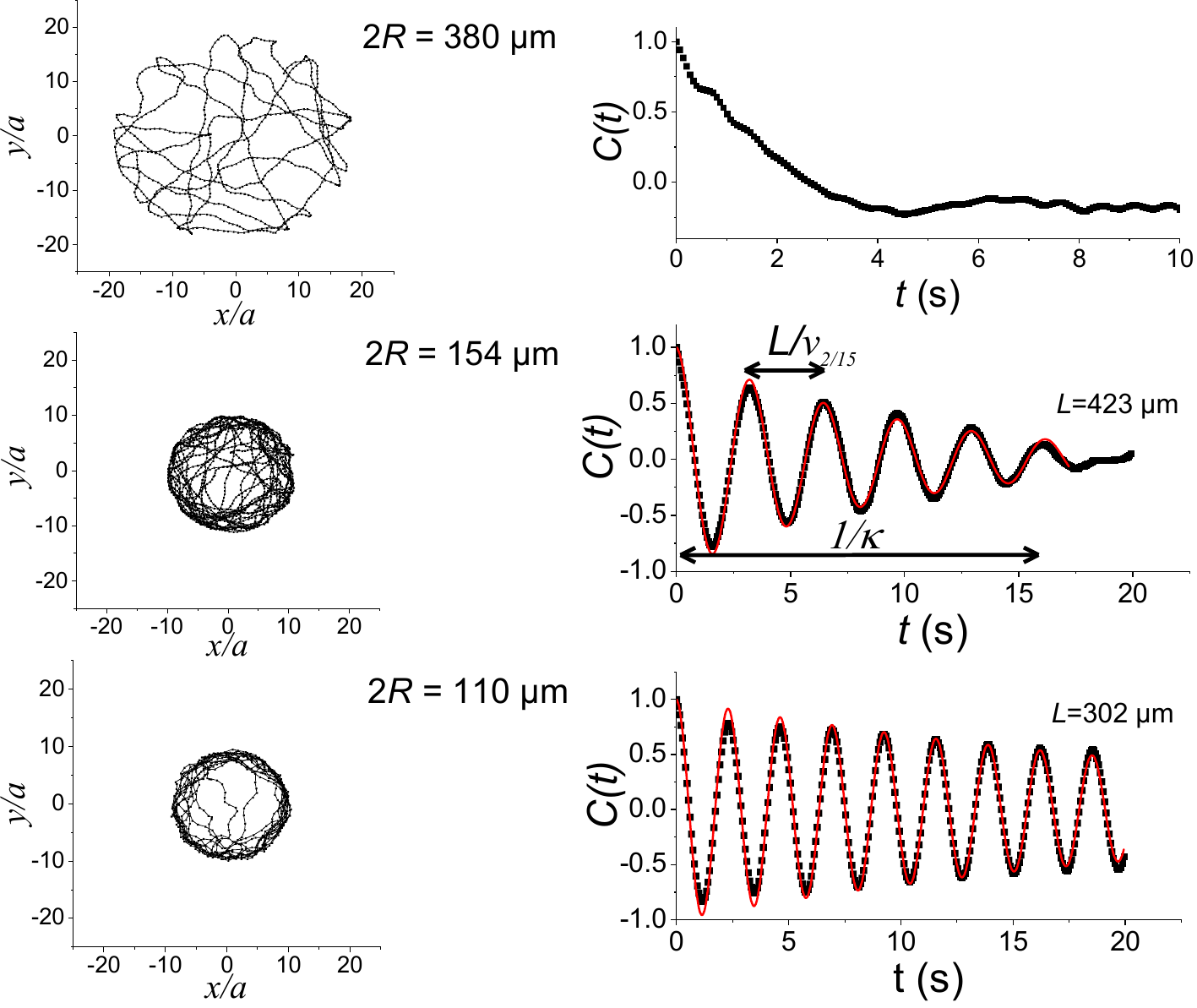}
\caption{\label{cavitytraj} Trajectories of individually trapped CR cells in circular cavities of varying diameter $2R$ (380 µm, 154 µm, and 110 µm), recorded over 1 to 2 min. Each trajectory is accompanied by its corresponding directional autocorrelation function $C(t)$. For cavities where $2R < L_0$, the autocorrelator is fitted to the damped sinusoid from eq. \ref{equa1} with the swimming speed fixed at $v = v_{2/15}=130$ µm/s, yielding the path length traveled per period $L$ and the decay time $1/\kappa$.}
\end{figure}

We study the swimming dynamics of individual CR cells trapped in isolated circular PDMS cavities of height 70 µm and varying diameter $2R$, ranging from 110 µm to 436 µm. Single-cell occupancy is enforced by controlled dilution of the suspension, following the protocol of \cite{ostapenko2018curvature}. As depicted in fig. \ref{cavitytraj}, cell trajectories reveal a clear confinement-dependent transition in swimming behavior. For large cavities ($2R > L_0$), cells exhibit bulk active Brownian dynamics similar to those observed in bulk, exploring the entire cavity area with occasional wall contacts. As the diameter decreases below $L_0$, cells progressively transition toward quasi-circular wall-following trajectories, in which the swimmer persistently grazes the curved boundary with rare excursions toward the center of the cavity. This transition mirrors that reported for active Brownian Janus particles in circular traps \cite{volpe2011microswimmers}, where an equivalent phenomenology arises purely from steric confinement, without invoking hydrodynamic interactions: upon collision with a concave wall, a cell must sample an increasingly wide range of orientations before finding a direction that points away from the surface, generating an effective geometric trapping whose strength grows with wall curvature.

\begin{figure*}[t] % [t] est fortement recommandé pour les figures double-colonne
    \centering
    
    % Première sous-figure
    \begin{subfigure}{0.48\textwidth} % Environ 50% de la largeur du texte
        \centering
        \includegraphics[width=\textwidth]{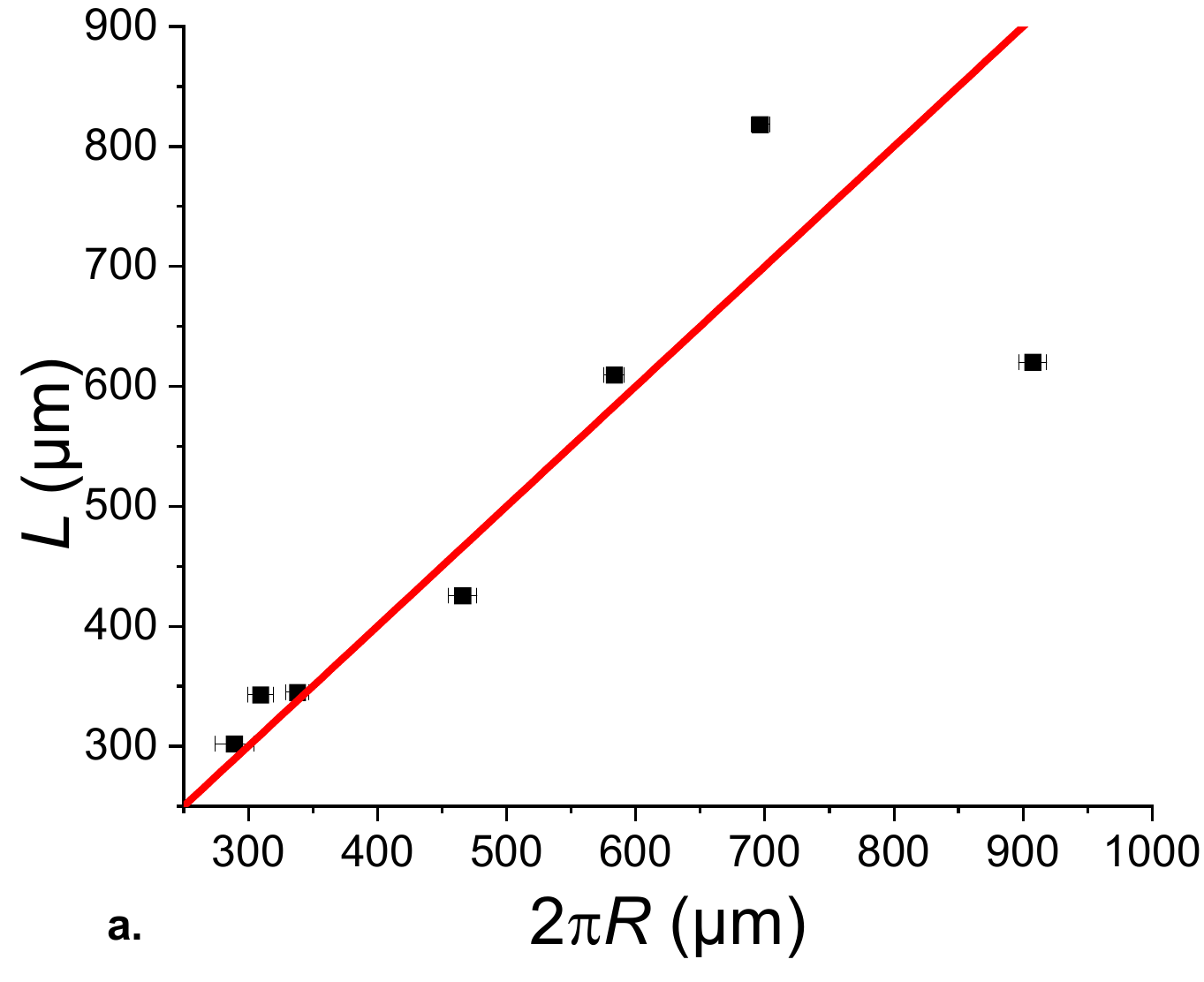}
    \end{subfigure}
    \hfill % Ajoute un espace flexible entre les deux images
    % Deuxième sous-figure
    \begin{subfigure}{0.48\textwidth}
        \centering
        \includegraphics[width=\textwidth]{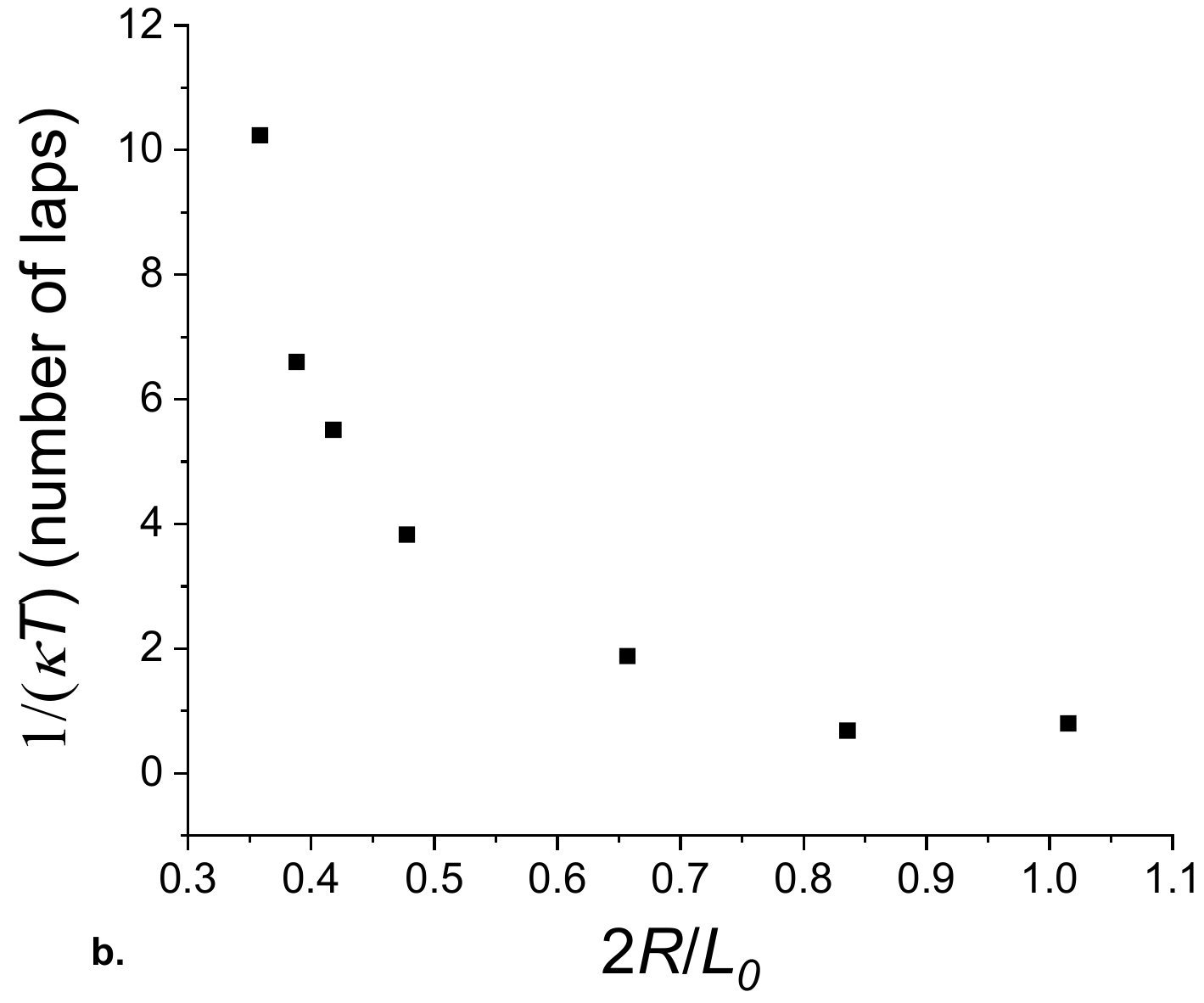}
    \end{subfigure}

    \caption{a. Path length per lap $L$ as a function of cavity perimeter $2\pi R$. Experimental data follow a linear trend marked by the first bisector drawn in red, confirming statistical wall-following for $2R < L_0$. b. Periodical orientational persistence time $\kappa$, expressed in number of laps $1/(\kappa T)$, as a function of normalized cavity diameter $2R/L_0$.}
    \label{L}
\end{figure*}

These two dynamical regimes are clearly reflected in the directional autocorrelation function $C(t)$. For large cavities, $C(t)$ decays exponentially, consistent with persistent active Brownian statistics in bulk. For cavities with $2R < L_0$, the periodic return of the swimmer to the same orientation after each lap around the wall gives rise to an oscillatory $C(t)$, well described by a damped sinusoid:
\begin{equation} \label{equa1}
    C(t) = \cos(2\pi v t/L) \exp(-\kappa t) ,
\end{equation}
where $L$ is the length of the path per lap, $v$ the mean swimming speed, and $\kappa$ the orientational decay rate. Fitting this expression to experimental autocorrelators, we extract $L$ as a function of the perimeter of the cavity $2\pi R$. As shown in fig. \ref{L}a, for all cavities with $2R < L_0$, $L \approx 2 \pi R$ confirming that the swimmer statistically follows the circular wall over this diameter range, with negligible bulk excursions. The damping coefficient $\kappa$ further quantifies how many laps the cell completes before losing directional memory. As shown in fig. \ref{L}b. The duration $1/\kappa$ far exceeds the intrinsic tumbling time scale $t_0$ for small cavities, and decreases monotonically as $2R$ approaches $L_0$, at which point a single lap suffices to fully decorrelate the swimming direction. This result demonstrates that geometric confinement in circular cavities generates a new form of motility persistence, externally imposed by the environment and tunable through cavity size, with a characteristic timescale substantially exceeding the swimmer's intrinsic reorientation dynamics.

These results establish circular confinement as a purely geometric source of enhanced persistence of motility. We now extend this framework to a dumbbell, geometry in which two of such reservoirs are connected by a straight channel, allowing us to probe how escape dynamics emerge from the interplay between cavity trapping and channel rectification. Two circular reservoirs of diameter $2R$ are connected by a variable width straight channel $d$, combining the confinement effects identified separately in channels and cavities. Individual cell trajectories (fig. \ref{merge}c) confirm that cells are not permanently trapped in a single compartment but alternate between reservoirs by traversing the connecting channel, showing wall-following in the cavities, and rectified swimming within the channel.

\begin{figure}[h]
\includegraphics[width=\columnwidth]{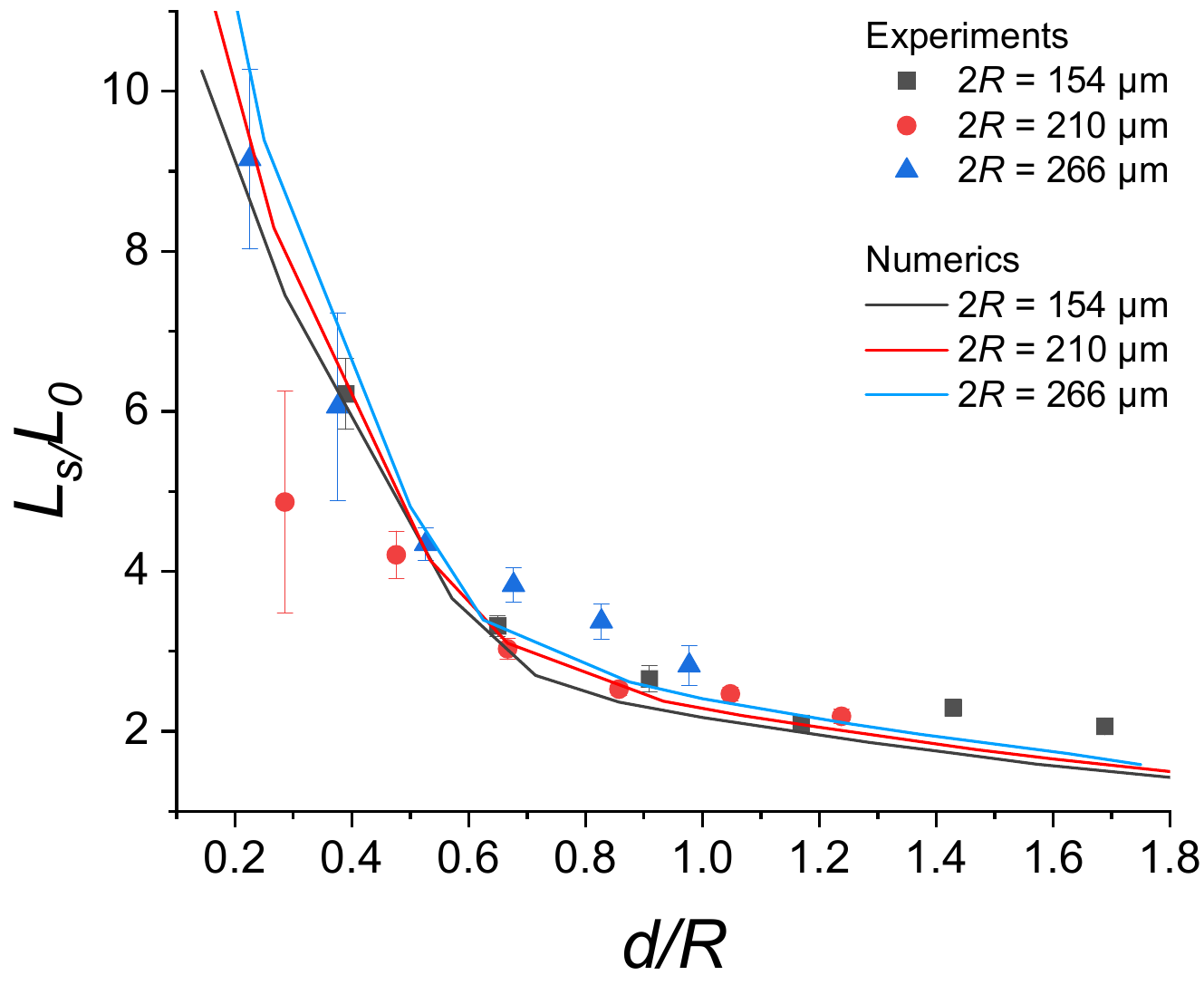}
\caption{\label{travel} Mean dwell length  of a microswimmer trapped individually normalized as $L_s/L_0$ in one cavity before crossing the connecting channel, as a function of channel width normalized by the cavirty radius $d/R$ and for several cavity diameters.}
\end{figure}

 To quantify inter-compartment dynamics, we measure the dwell length $L_s$—the mean length of the path traveled within a reservoir before escape—as a function of $d$ for several values of $R$. In Fig. \ref{travel}, as expected, $L_s$ decreases with increasing $d$ at fixed $R$, reflecting the higher geometric probability of encountering a wider exit. When $L_s$ is plotted against the normalized opening $d/R$, the data from cavities of different radii $R$ collapse into a single master curve, indicating that the escape dynamics are governed by the ratio $d/R$ alone.

 To assess whether this escape dynamics can be captured by a purely stochastic swimming model, we applied the numerical simulations of Active Brownian Particles in the dumbbell geometry. In our numerical simulations, the values of $L_s$ are in qualitative agreement with the experimental data across the range of $d/R$ explored, suggesting that the inter-compartment transfer dynamics are predominantly governed by geometric confinement rather than by swimmer-wall hydrodynamic coupling. The overall consistency between the two approaches indicates that no clear hydrodynamic signature seems to be required to account for the observed escape behavior, reinforcing the picture of a geometry-driven and probability-controlled transfer process. Signatures of hydrodynamic interactions may be more clearly evidenced in experiments performed under stronger confinement, such as those conducted in channel geometries.

\section{\label{discussion}Discussion and Conclusion}
A central outcome of this study is the identification of the respective roles of steric contact and hydrodynamic interactions across confinement geometries. In straight channels and circular cavities of moderate size comparable to or larger than $L_0$, near-wall accumulation and orientational biases are quantitatively reproduced by Active Brownian Particle simulations treating the swimmer as a point particle subject to purely steric boundary conditions, demonstrating that geometric trapping and wall-induced reorientation do not require hydrodynamic interactions to emerge. However, this steric picture progressively breaks down as confinement is increased: the non-monotonic velocity enhancement observed in strongly confined channels exceeds steric predictions, which is consistent with a hydrodynamic coupling between the swimmer's flow field and nearby boundaries. These observations suggest that steric contributions dominate in moderate confinement, while hydrodynamic effects become significant as the characteristic geometric length scale falls well below $L_0$. The confinement ratio therefore emerges as the key parameter controlling the crossover between these two regimes.

Beyond interaction mechanisms, our results demonstrate that environmental geometry alone constitutes a powerful passive means of controlling the motility of microswimmer. In circular cavities with $2R < L_0$, the swimming regime transitions from diffusive active Brownian exploration to quasi-circular wall-following, generating a persistent geometry-induced state whose decay time $1/\kappa$ far exceeds the intrinsic tumbling timescale $t_0$ — without any external forcing, chemical gradient, or light stimulus \cite{berg1972chemotaxis, brun2020deflection}. This places our findings in a broader context of passive motility control. The robustness of this mechanism is underscored by its independence from the swimmer's detailed locomotion strategy: analogous transitions have been reported for synthetic Janus particles \cite{volpe2014simulation} and for CR in elliptic cavities \cite{ostapenko2018curvature}, both reproduced by active Brownian simulations of a point-like particle without hydrodynamics. Nevertheless, systematic discrepancies between point-particle simulations and experimental density profiles—most notably the offset in near-wall accumulation peaks in channel confinement—highlight the limitations of neglecting the finite size of the cell body and flagella. Extensions of the present framework incorporating finite swimmer size, flagellar dynamics, and far-field hydrodynamic interactions would be required to achieve quantitative agreement across all confinement regimes.

The elementary geometries investigated here—channels, circular cavities, and dumbbell topologies—provide a controlled set of building blocks for understanding microswimmer-boundary interactions, yet they remain far removed from the disordered, multiply-connected environments encountered in natural habitats such as sediments, soils, or coral structures. We previously established anomalous diffusion of CR in a periodic lattice of cylindrical obstacles \cite{brun2019effective, brun2020deflection}, deciphering a confinement-dependent enhancement of swimmer diffusivity in such a complex medium. The present study offers a mechanistic foundation for this phenomenology: by systematically isolating the contributions of channel-like confinement, curved wall interactions, and inter-compartment transport in elementary geometries, we identify the elementary processes—orientational rectification, near-wall accumulation, geometry-induced persistence, and hydrodynamic speed modulation—whose interplay is expected to underlie the emergent transport behavior observed in the pillar network. Our results thus provide the missing picture behind the macroscopic diffusive phenomenology reported in that study and lay the groundwork for a quantitative, geometry-informed model of microswimmer transport in porous and structured media.

\bibliography{main}% Produces the bibliography via BibTeX.

@article{lauga2009hydrodynamics,
  title={The hydrodynamics of swimming microorganisms},
  author={Lauga, Eric and Powers, Thomas R.},
  journal={Reports on Progress in Physics},
  volume={72},
  number={9},
  pages={096601},
  year={2009},
  publisher={IOP Publishing}
}

@article{guasto2012fluid,
  title={Fluid mechanics of planktonic microorganisms},
  author={Guasto, Jeffrey S. and Rusconi, Roberto and Stocker, Roman},
  journal={Annual Review of Fluid Mechanics},
  volume={44},
  pages={373--400},
  year={2012},
  publisher={Annual Reviews}
}

@article{altshuler2013flow,
  title={Flow-controlled densification and anomalous dispersion of \textit{E. coli} through a constriction},
  author={Altshuler, Ernesto and Mi{\~n}o, Guillermo and P{\'e}rez-Penichet, Carlos and del R{\'i}o, Lev and Lindner, Anke and Rousselet, Annie and Cl{\'e}ment, Eric},
  journal={Soft Matter},
  volume={9},
  number={6},
  pages={1864--1870},
  year={2013},
  publisher={Royal Society of Chemistry}
}

@article{berdakin2013influence,
  title={Influence of swimming strategy on microorganism separation by asymmetric obstacles},
  author={Berdakin, Iv{\'a}n and Jeyaram, Yogesh and Moshchalkov, Victor V. and Venken, Luc and Dierckx, Sam and Vanderleyden, Jos and Silhanek, Alejandro V.},
  journal={Physical Review E},
  volume={87},
  number={5},
  pages={052702},
  year={2013},
  publisher={APS}
}

@article{berg1972chemotaxis,
  title={Chemotaxis in \textit{Escherichia coli} analysed by three-dimensional tracking},
  author={Berg, Howard C. and Brown, Douglas A.},
  journal={Nature},
  volume={239},
  number={5374},
  pages={500--504},
  year={1972},
  publisher={Nature Publishing Group}
}

@article{garcia2011random,
  title={Random walk of a swimmer in a low-Reynolds-number medium},
  author={Garcia, Michael and Berti, Stefano and Peyla, Philippe and Rafa{\"i}, Salima},
  journal={Physical Review E},
  volume={83},
  number={3},
  pages={035301},
  year={2011},
  publisher={APS}
}

@article{mino2011enhanced,
  title={Enhanced diffusion due to active swimmers at a solid surface},
  author={Mi{\~n}o, Guillermo and Mallouk, Thomas E. and Darnige, Thierry and Hoyos, Mauricio and Dauchet, Josselin and Dunstan, Jocelyn and Cl{\'e}ment, Eric},
  journal={Physical Review Letters},
  volume={106},
  number={4},
  pages={048102},
  year={2011},
  publisher={APS}
}

@article{berke2008hydrodynamic,
  title={Hydrodynamic attraction of swimming microorganisms by surfaces},
  author={Berke, Allison P. and Turner, Linda and Berg, Howard C. and Lauga, Eric},
  journal={Physical Review Letters},
  volume={101},
  number={3},
  pages={038102},
  year={2008},
  publisher={APS}
}

@article{spagnolie2012hydrodynamics,
  title={Hydrodynamics of self-propulsion near a boundary: predictions and accuracy of far-field approximations},
  author={Spagnolie, Saverio E. and Lauga, Eric},
  journal={Journal of Fluid Mechanics},
  volume={700},
  pages={105--147},
  year={2012},
  publisher={Cambridge University Press}
}

@article{kantsler2013ciliary,
  title={Ciliary contact interactions dominate surface scattering of swimming eukaryotes},
  author={Kantsler, Vasily and Dunkel, J{\"o}rn and Polin, Marco and Goldstein, Raymond E.},
  journal={Proceedings of the National Academy of Sciences},
  volume={110},
  number={4},
  pages={1187--1192},
  year={2013},
  publisher={National Academy of Sciences}
}

@article{brun2019effective,
  title={Effective diffusivity of microswimmers in a crowded environment},
  author={Brun-Cosme-Bruny, Marvin and Bertin, Eric and Coasne, Benoit and Peyla, Philippe and Rafa{\"i}, Salima},
  journal={The Journal of Chemical Physics},
  volume={150},
  number={10},
  year={2019},
  publisher={AIP Publishing}
}

@article{thery2021rebound,
  title={Rebound and scattering of motile \textit{Chlamydomonas} algae in confined chambers},
  author={Th{\'e}ry, Antoine and Wang, Yibo and Dvoriashyna, Mariia and Eloy, Christophe and Elias, Florence and Lauga, Eric},
  journal={Soft Matter},
  volume={17},
  number={18},
  pages={4857--4873},
  year={2021},
  publisher={Royal Society of Chemistry}
}

@article{ao2014active,
  title={Active Brownian motion in a narrow channel},
  author={Ao, Xin and Ghosh, Pulak K. and Li, Yunyun and Schmid, Gerhard and H{\"a}nggi, Peter and Marchesoni, Fabio},
  journal={The European Physical Journal Special Topics},
  volume={223},
  number={14},
  pages={3227--3242},
  year={2014},
  publisher={Springer}
}

@article{volpe2014simulation,
  title={Simulation of the active Brownian motion of a microswimmer},
  author={Volpe, Giovanni and Gigan, Sylvain and Volpe, Giorgio},
  journal={American Journal of Physics},
  volume={82},
  number={7},
  pages={659--664},
  year={2014},
  publisher={American Association of Physics Teachers}
}

@article{ostapenko2018curvature,
  title={Curvature-guided motility of microalgae in geometric confinement},
  author={Ostapenko, Taras and Schwarzendahl, Fabian Jan and B{\"o}ddeker, Thomas J. and Kreis, Christian T. and Cammann, Jan and Mazza, Marco G. and B{\"a}umchen, Oliver},
  journal={Physical Review Letters},
  volume={120},
  number={6},
  pages={068002},
  year={2018},
  publisher={APS}
}

@article{brun2020deflection,
  title={Deflection of phototactic microswimmers through obstacle arrays},
  author={Brun-Cosme-Bruny, Marvin and F{\"o}rtsch, Andre and Zimmermann, Walter and Bertin, Eric and Peyla, Philippe and Rafa{\"i}, Salima},
  journal={Physical Review Fluids},
  volume={5},
  number={9},
  pages={093302},
  year={2020},
  publisher={APS}
}

@article{chepizhko2019ideal,
  title={Ideal circle microswimmers in crowded media},
  author={Chepizhko, Oleksandr and Franosch, Thomas},
  journal={Soft Matter},
  volume={15},
  number={3},
  pages={452--461},
  year={2019},
  publisher={Royal Society of Chemistry}
}

@article{alalam2022active,
  title={Active jamming of microswimmers at a bottleneck constriction},
  author={Al Alam, Elias and Brun-Cosme-Bruny, Marvin and Borne, Valentine and Faure, Simon and Maury, Bertrand and Peyla, Philippe and Rafa{\"i}, Salima},
  journal={Physical Review Fluids},
  volume={7},
  number={9},
  pages={L092301},
  year={2022},
  publisher={APS}
}

@article{parra2014casimir,
  title={Casimir effect in swimmer suspensions},
  author={Parra-Rojas, Crist{\'o}bal and Soto, Rodrigo},
  journal={arXiv preprint arXiv:1404.4857},
  year={2014}
}

@book{harris2009chlamydomonas,
  title={The \textit{Chlamydomonas} Sourcebook},
  author={Harris, Elizabeth H.},
  volume={1},
  pages={293--302},
  year={2009},
  editor={Stern, D. B. and Witman, G.},
  publisher={Elsevier},
  address={San Diego, CA}
}

@article{polin2009chlamydomonas,
  title={\textit{Chlamydomonas} swims with two ``gears'' in a eukaryotic version of run-and-tumble locomotion},
  author={Polin, Marco and Tuval, Idan and Drescher, Knut and Gollub, Jerry P. and Goldstein, Raymond E.},
  journal={Science},
  volume={325},
  number={5939},
  pages={487--490},
  year={2009},
  publisher={AAAS}
}

@article{qin2010soft,
  title={Soft lithography for micro- and nanoscale patterning},
  author={Qin, Dong and Xia, Younan and Whitesides, George M.},
  journal={Nature Protocols},
  volume={5},
  number={3},
  pages={491--502},
  year={2010},
  publisher={Nature Publishing Group}
}

@article{garcia2013light,
  title={Light control of the flow of phototactic microswimmer suspensions},
  author={Garcia, Xavier and Rafa{\"i}, Salima and Peyla, Philippe},
  journal={Physical Review Letters},
  volume={110},
  number={13},
  pages={138106},
  year={2013},
  publisher={APS}
}

@article{allan2018trackpy,
  title={trackpy: Trackpy v0.4.1},
  author={Allan, Daniel B. and Caswell, Thomas and Keim, Nathan C. and van der Wel, Casper M.},
  journal={Zenodo},
  year={2018}
}

@article{crocker1996methods,
  title={Methods of digital video microscopy for colloidal studies},
  author={Crocker, John C. and Grier, David G.},
  journal={Journal of Colloid and Interface Science},
  volume={179},
  number={1},
  pages={298--310},
  year={1996},
  publisher={Elsevier}
}

@article{ishikawa2025transport,
  title={Transport phenomena in microswimmer suspensions: migration, collective motion, diffusion and rheology},
  author={Ishikawa, Takuji},
  journal={Journal of Fluid Mechanics},
  volume={1016},
  pages={P1},
  year={2025},
  publisher={Cambridge University Press}
}

@article{wu2016amoeboid,
  title={Amoeboid swimming in a channel},
  author={Wu, Hao and Farutin, Alexander and Hu, Wei-Fan and Thi{\'e}baud, Marine and Rafa{\"\i}, Salima and Peyla, Philippe and Lai, Ming-Chih and Misbah, Chaouqi},
  journal={Soft matter},
  volume={12},
  number={36},
  pages={7470--7484},
  year={2016},
  publisher={Royal Society of Chemistry}
}

@article{volpe2011microswimmers,
  title={Microswimmers in patterned environments},
  author={Volpe, Giovanni and Buttinoni, Ivo and Vogt, Dominik and K{\"u}mmerer, Hans-J{\"u}rgen and Bechinger, Clemens},
  journal={Soft Matter},
  volume={7},
  number={19},
  pages={8810--8815},
  year={2011},
  publisher={Royal Society of Chemistry}
}

@article{lagoin2025enhanced,
  title={Enhanced dispersion of active microswimmers in confined flows},
  author={Lagoin, Marc and Lacherez, Juliette and de Tournemire, Guirec and Badr, Ahmad and Amarouchene, Yacine and Allard, Antoine and Salez, Thomas},
  journal={Proceedings of the National Academy of Sciences},
  volume={122},
  number={50},
  pages={e2519691122},
  year={2025},
  publisher={National Academy of Sciences}
}

@article{bardfalvy2024collective,
  title={Collective motion in a sheet of microswimmers},
  author={B{\'a}rdfalvy, D{\'o}ra and {\v{S}}kult{\'e}ty, Viktor and Nardini, Cesare and Morozov, Alexander and Stenhammar, Joakim},
  journal={Communications Physics},
  volume={7},
  number={1},
  pages={93},
  year={2024},
  publisher={Nature Publishing Group UK London}
}

\end{document}